\documentclass[11pt,twoside]{article}

\usepackage{asp2006}
\usepackage{epsf}
\usepackage{psfig}
\usepackage{lscape}

\begin{document}
\title{The Advection of Supergranules by Large-Scale Flows}
\author{David H. Hathaway}
\affil{NASA Marshall Space Flight Center, Huntsville, AL 35812 USA}

\author{Peter E. Williams, Manfred Cuntz}
\affil{Department of Physics, University of Texas at Arlington, Arlington, TX 76019 USA}

\begin{abstract}
We produce a 10-day series of simulated Doppler images at a 15-minute cadence that
reproduces the spatial and temporal characteristics seen in the SOHO/MDI Doppler data.
Our simulated data contains a spectrum of cellular flows with but two necessary components ---
a granule component that peaks at wavenumbers of about 4000 and a supergranule component
that peaks at wavenumbers of about 110. We include the advection of these cellular
components by a differential rotation profile that depends on latitude and wavenumber (depth).
We further mimic the evolution of the cellular pattern by introducing random variations to
the amplitudes and phases of the spectral components at rates that reproduce the level
of cross-correlation as a function of time and latitude. Our simulated data do not
include any wave-like characteristics for the supergranules yet can accurately reproduce
the rotation characteristics previously attributed to wave-like characteristics.
\end{abstract}

\section{Introduction}

Supergranules are cellular flow structures observed in the solar photosphere with typical
diameters of about 30 Mm and lifetimes of about one day. They cover the entire surface
of the Sun and are intimately involved with the structure and evolution of the magnetic
field in the photosphere. The magnetic structures of the chromospheric network form at the boundaries of
these cells and magnetic elements are shuffled about the surface as the cells evolve.
The diffusion of magnetic elements by the evolving supergranules has long been associated
with the evolution of the Sun's magnetic field [\cite{Leighton64}].

Supergranules were discovered by \cite{Hart54}. While these cellular flows
were quickly identified as convective features [\cite{LeightonNoyesSimon62}]
the difficulty of detecting any associated thermal features consistent with that
dentification (i.e. hot cell centers) has made this identification somewhat
problematic [\cite{Worden75}].

The rotation of the supergranules has added further mystery to their nature.
\citet{Duvall80} cross-correlated the equatorial Doppler velocity patterns
and found that the supergranules rotated more rapidly than the plasma
at the photosphere and that even faster rotation rates were obtained when longer
(24-hour vs. 8-hour) time intervals were used. He attributed this behavior to a
surface shear layer [proposed by \cite{FoukalJokipii75} and \cite{Foukal79} and
modeled by \cite{GilmanFoukal79}] in which larger, longer-lived, cells extend deeper
into the more rapidly rotating layers. \cite{SnodgrassUlrich90} used
data from Mount Wilson Observatory to find the rotation rate at different
latitudes and noted that the rotation rates for the Doppler pattern were some 4\%
faster than the spectroscopic rate and, more interestingly, some 2\% faster than the
magnetic features and sunspots.
More recently \cite{BeckSchou00} used a 2D Fourier transform
method to find that the Doppler pattern rotates more rapidly than the shear layer itself
and that larger features do rotate more rapidly than the smaller features. They suggested
that supergranules have wave-like characteristics with a preference for prograde propagation.

In a previous paper [\cite{Hathaway_etal06}] we showed that this ``super-rotation'' of
the Doppler pattern could be attributed to projection effects associated with the
Doppler signal itself. As the velocity pattern rotates across the field of view its
line-of-sight component is modulated in a way that essentially adds another half wave
and gives a higher rotation rate that is a function of wavenumber. In that paper
we took a fixed velocity pattern (which had spatial characteristics that matched the
SOHO/MDI data) and rotated it rigidly to show this ``super-rotation'' effect.
While this indicated that this Doppler projection effect should be accounted for,
the fixed pattern could not account for all the variations reported by
\cite{BeckSchou00}. Furthermore, when \cite{Schou03} ``divided-out'' the line-of-sight
modulation he still saw prograde and retrograde moving components.

In this paper we report on our analyses of simulated data in which the supergranules
are advected by a differential rotation that varies with both latitude and depth.
The data is designed to faithfully mimic the SOHO/MDI data that was analyzed in
\cite{BeckSchou00} and \cite{Schou03} and the analyses are reproductions of those
done in earlier studies. 

\section{The Data}

The full-disk Doppler images from SOHO/MDI [\cite{Scherrer_etal95}]
are obtained at a 1-minute
cadence to resolve the temporal variations associated with the p-mode oscillations.
We [c.f. \cite{Hathaway_etal00} and \cite{BeckSchou00}] have temporally filtered the
images to remove the p-mode signal by using a 31-minute long tapered Gaussian with a FWHM
of 16 minutes on sets of 31 images that were de-rotated to register each to the central image.
Series of these filtered images were formed at a 15-minute cadence over the 60-day
MDI Dynamics Runs in 1996 and 1997. This filtering process effectively removes the p-mode
signal and leaves behind the Doppler signal from flows with temporal variations
longer than about 16 minutes.
Supergranules, with typical wavenumbers of about 110, are very well
resolved in this data (at disk center wavenumbers of about 1500 are resolved). While
granules are not well resolved, they do appear in the data as pixel-to-pixel and
image-to-image ``noise,'' as a convective blue shift (due to the correlation between
brightness and updrafts), and as resolved structures for the largest members.

The simulated data are constructed in the manner described in \cite{Hathaway88},
\cite{Hathaway92}, \cite{Hathaway_etal00}, and \cite{Hathaway_etal02}
from vector velocities generated by an input spectrum of complex
spectral coefficients for the radial, poloidal, and toroidal components.

To simulate the observed line-of-sight velocity the three vector velocity
components are calculated
on a grid with 1024 points in latitude and 2048 points in longitude.
A Doppler velocity image is constructed by determining the longitude and
latitude at a point on the image, finding the vector velocity at that point
using bi-cubic interpolation, and then projecting that vector velocity
onto the line-of-sight. The line-of-sight velocities at
an array of 16 points within each pixel are determined and the average
taken to simulate the integration over a pixel in the acquisition of
the actual MDI Doppler data.

With the current simulations we have added two changes that were not included in our
previous work on individual Doppler images. First, we add velocity ``noise'' at each
pixel. This represents the contribution from the spatially unresolved granules that,
nonetheless, have temporal varibility that is not filtered out by the 31-minute
temporal filter. This noise has a center-to-limb variation due to the foreshortened
area covered by each pixel and is randomly varied from pixel to pixel and from one
Doppler image to the next. The noise level is determined by matching the initial drop in
correlation from one image to the next that is seen in the MDI data.
Secondly, we treat the instrumental blurring in a more
realistic manner. Previously we took the Doppler velocity image and convolved it with
an MDI point-spread-function. We now make red and blue intensity images from our
Doppler velocity image and a simple limb darkened intensity image, convolve those with
an MDI point-spread-function, and construct a blurred Doppler velocity image from the
difference divided by the sum.
This process yields a Doppler velocity image that is virtually indistinguishable from
an MDI Doppler velocity image.

The velocity pattern is evolved in time by introducing changes to the spectral coefficients
based on two processes - the advection by an axisymmetric zonal flow (differential
rotation) and random processes that lead to the finite lifetime of the cells.

The advection by the differential rotation is governed by an advection equation

\begin{equation}
{\partial u \over \partial t} = - \Omega(\theta) {\partial u \over \partial \phi}
\end{equation}

\noindent where $u$ is a vector velocity component and $\Omega(\theta)$
is the differential rotation profile.
We represent $u$ as a series of spherical harmonic components and
project this advection equation onto a single spherical harmonic which gives a series
of coupled equations for the evolution of the spectral coefficients. Solid body
rotation simply introduces a phase variation for each coefficient. Differential
rotation couples the change in one spectral coefficient to spectral coefficients
with wavenumbers $\ell \pm 2$ and $\ell \pm 4$ for differential rotation dependent on
$\cos^2 \theta$ and $\cos^4 \theta$.

The finite lifetimes for the cells are simulated by introducing random perturbations
to the spectral coefficient amplitudes and phases. The size of these perturbations
increases with wavenumber to give shorter lifetimes to smaller cells.

\section{The Analyses}

Several anaylsis programs were applied to both the MDI data and the simulated data.
Convection spectra for individual images were obtained using the methods described
by \cite{Hathaway87} and \cite{Hathaway92} --- the Doppler signal due to the motion
of the observer is removed, the convective blue shift signal is identified and removed,
the data is mapped to heliographic coordinates, the axisymmetric flow signals due
to differential rotation and meridional circulation are identified and removed,
and the remaining signal is projected onto spherical harmonics. The averaged spectra
from the 1996 MDI Dynamics Run and from our 10-day simulated data run are
nearly perfectly matched at all wavenumbers.
This match is obtained by adjusting the input spectrum for the simulated data.
This spectrum contains two Lorentzian-like
spectral components --- a supergranule component centered on $\ell \sim 110$ with
a width of about 100 and a granule component centered on $\ell \sim 4000$ with a
width of about 4000. The MDI spectrum is well matched with just these two components
without the addition of a mesogranule component [\cite{November_etal81}].
In fact, we find a distinct {\em dip} in the spectrum at wavenumbers $\ell \sim 500$
that should be representative of mesogranules. This dip is also seen in spectra of
the MDI high resolution data [\cite{Hathaway_etal00}].

Additional analyses are applied to the data after it has been mapped onto heliographic
coordinates. Longitudal strips of this data centered on latitudes from $75^\circ$ south
to $75^\circ$ north were cross-correlated with corresponding strips
from later images as was done by \citet{Duvall80} and \citet{SnodgrassUlrich90}.
The longitudinal shift of the cross-correlation peak gives the rotation rate while
the height of this peak is associated with cell lifetimes. These strips were also
Fourier analyzed in longitude to get spectral coefficients and those coefficents
were Fourier anaylzed in time over 10-day intervals as was done by \citet{BeckSchou00}
to get rotation rates as functions of wavenumber.

Fig. 1 shows the rotation rates from the cross-correlation analysis.
The rotation profiles from the simulated data match those from the MDI data.
Both show faster rotation rates for longer time lags as noted by \citet{Duvall80}
and by \citet{SnodgrassUlrich90}. This indicates that we have found the right
latitudinal differential rotation profile.

\begin{figure}[!ht]
\plotone{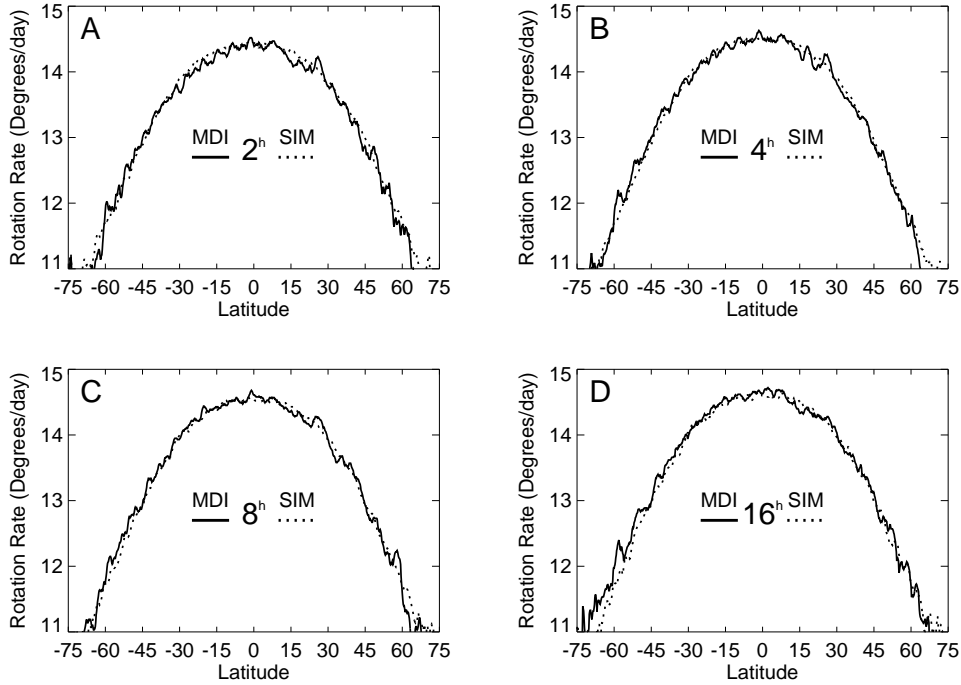}
\caption{
Rotation profiles from cross-correlation analyses of MDI data (solid lines) and simulated
data (dotted lines). All profiles match at virtually all latitudes and time lags
(Panel A --- 2-hour, Panel B --- 4-hour, Panel C --- 8-hour, Panel D --- 16-hour).
Note that the measured equatorial rotation rate increases with time lag for
both datasets.
}
\end{figure}

The strength of the correlations as functions of latitude and time lag
for both the MDI data and the simulated data are also well matched.
This indicates that we have found the right lifetimes for the cells.

We have also reproduced the analysis of \cite{Schou03}. The data strips are apodized
and multiplied by longitude dependent functions designed to remove the Doppler
projection effect and to isolate either longitudinal motions or latitudinal motions.
The strips are shifted in longitude according to the differential rotation rate and then
Fourier analysed in space and time to obtain ``$k\omega$'' diagrams.

Fig. 2A shows the equatorial rotation rate as a function of wavenumber for
the simulated data while Fig. 2B shows the $k\omega$ diagram.
These should be compared to Fig. 4 of \cite{BeckSchou00}
and Fig. 8 of \cite{Schou03} respectively.

\begin{figure}[!ht]
\plotone{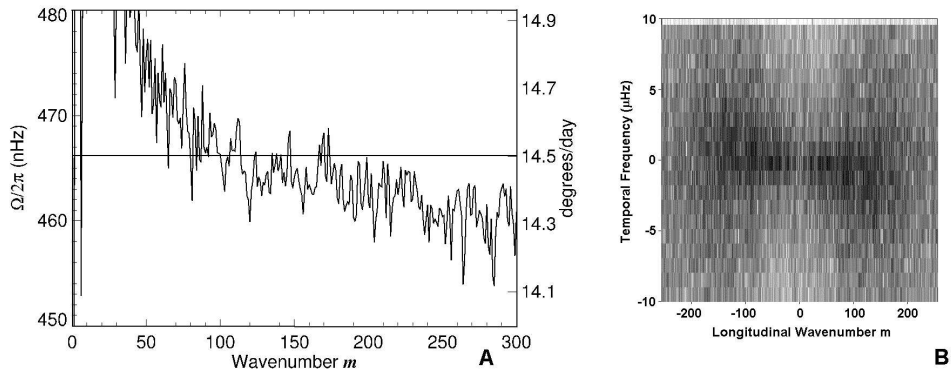}
\caption{
Panel A. The equatorial rotation rate as a function of wavenumber from our analysis
of the simulated data showing the increase at small wavenumbers.
Most of this increase is due to the Doppler velocity projection.
Panel B. The $k\omega$ diagram from our analysis of the simulated data showing
both prograde and retrograde moving elements.
This spread in power is due to the random evolution of the cellular pattern.
}
\end{figure}

\section{Conclusions}

We have produced simulated data in which the cellular structures (supergranules) are advected
by differential rotation and evolve by uncorrelated random changes. When we compare results
from analyses of this data with those from analyses of the MDI data we find that the
simulated data exhibits the same characteristics as the MDI data --- the visual structures,
the power spectra, the rotation characteristics, and the evolution rates all match.
While some of these characteristics have been attributed to wave-like properties
[c.f. \cite{BeckSchou00} and \cite{Schou03}] our simulated data is simply advected
by a zonal flow (differential rotation) with speeds that never exceed those determined from
helioseismology [\cite{Schou_etal98}]. The differential rotation we impose does, however,
have a dependence on wavenumber $\ell$. If we assume that the rotation rate of cells with
diameters, $D = 2 \pi R_\odot/\ell$, reflects the rotation rate at a depth, $d = D/2$,
then the surface shear layer indicated by our differential rotation has a thickness
of about 20 Mm --- somewhat thinner than the 30 Mm suggested by helioseismic inversions
[\cite{Schou_etal98}].

\acknowledgements
We would like to thank NASA for its support of this research through a grant
from the Heliophysics Guest Investigator Program to NASA Marshall Space Flight Center
and The University of Texas Arlington. We would also like to thank the SOHO/MDI team
for the critical role they played in producing the raw MDI data and John Beck in particular
for implementing the temporal averaging of that data to remove the p-mode noise.


\begin{thebibliography}{}
\bibitem[\protect\citeauthoryear{Beck \& Schou}{2000}]{BeckSchou00}
  Beck, J. G., \& Schou, J. 2000, \solphys, 193, 333

\bibitem[\protect\citeauthoryear{Duvall}{1980}]{Duvall80}
  Duvall Jr., T. L. 1980, \solphys, 66, 213

\bibitem[\protect\citeauthoryear{Foukal}{1979}]{Foukal79}
  Foukal, P. 1979, \apj, 218, 539

\bibitem[\protect\citeauthoryear{Foukal \& Jokipii}{1975}]{FoukalJokipii75}
  Foukal, P., \& Jokipii, R. 1975, \apjl, 199, L71

\bibitem[\protect\citeauthoryear{Gilman \& Foukal}{1979}]{GilmanFoukal79}
  Gilman, P. A., \& Foukal, P. 1979, \apj, 229, 1179

\bibitem[\protect\citeauthoryear{Hart}{1954}]{Hart54}
  Hart, A. B. 1954, \mnras, 114, 17

\bibitem[\protect\citeauthoryear{Hathaway}{1987}]{Hathaway87}
  Hathaway, D. H. 1987, \solphys, 108, 1

\bibitem[\protect\citeauthoryear{Hathaway}{1988}]{Hathaway88}
  Hathaway, D. H. 1988, \solphys, 117, 329

\bibitem[\protect\citeauthoryear{Hathaway}{1992}]{Hathaway92}
  Hathaway, D. H. 1992, \solphys, 137, 15

\bibitem[\protect\citeauthoryear{Hathaway et al.}{2000}]{Hathaway_etal00}
  Hathaway, D. H., Beck, J. G., Bogart, R. S., Bachmann, K. T.,
  Khatri, G., Petitto, J. M., Han, S., \& Raymond, J. 2000, \solphys, 193, 299

\bibitem[\protect\citeauthoryear{Hathaway et al.}{2002}]{Hathaway_etal02}
  Hathaway, D. H., Beck, J. G., Han, S., \& Raymond, J. 2002,
  \solphys, 205, 25

\bibitem[\protect\citeauthoryear{Hathaway, Williams, \& Cuntz}{2006}]{Hathaway_etal06}
  Hathaway, D. H., Williams, P. E., \& Cuntz,M. 2006, \apj, 644, 598

\bibitem[\protect\citeauthoryear{Leighton}{1964}]{Leighton64}
  Leighton, R. B. 1964, \apj, 140, 1559

\bibitem[\protect\citeauthoryear{Leighton, Noyes, \& Simon}{1962}]{LeightonNoyesSimon62}
  Leighton, R. B., Noyes, R. W., \& Simon, G. W. 1962, \apj, 135, 474

\bibitem[\protect\citeauthoryear{November et al.}{1981}]{November_etal81}
  November, L. J., Toomre, J., Gebbie, K. B., and  Simon, G. W. 1981, \apjl, 245, L123

\bibitem[\protect\citeauthoryear{Scherrer et al.}{1995}]{Scherrer_etal95}
  Scherrer, P. H., Bogart, R. S., Bush, R. I., Hoeksema, J. T.,
  Kosovichev, A. G., Schou, J., Rosenberg, W., Springer, L., Tarbell, T. D.,
  Title, A., Wolfson, C. J., Zayer, I., and the MDI Engineering Team 1995,
  \solphys, 162, 129

\bibitem[\protect\citeauthoryear{Schou}{2003}]{Schou03}
  Schou, J. 2003, \apjl, 596, L259

\bibitem[\protect\citeauthoryear{Schou et al.}{1998}]{Schou_etal98}
  Schou, J., Antia, H. M., Basu, S., Bogart, R. S., Bush, R. I., Chitre, S. M.,
  Christensen-Dalsgaard, J., Di Mauro, M. P., Dziembowski, W. A., Eff-Darwich, A.,
  Gough, D. O., Haber, D. A., Hoeksema, J. T., Howe, R., Korzennik, S. G.,
  Kosovichev, A. G., Larsen, R. M., Pijpers, F. P., Scherrer, P. H., Sekii, T.,
  Tarbell, T. D., Title, A. M., Thompson, M. J., \& Toomre, J. 1998, \apj, 505, 390

\bibitem[\protect\citeauthoryear{Snodgrass \& Ulrich}{1990}]{SnodgrassUlrich90}
  Snodgrass, H. B., \& Ulrich, R. K. 1990, \apj, 351, 309

\bibitem[\protect\citeauthoryear{Worden}{1975}]{Worden75}
  Worden, S. P. 1975, \solphys, 45, 521

\end{thebibliography}
\end{document}